\documentclass[aps,prc,twocolumn,floatfix,superscriptaddress]{revtex4-2}
\usepackage{epsfig}
\usepackage{amsmath}
\usepackage{color}
\usepackage{placeins}
\usepackage{graphicx}
\usepackage{soul}
\usepackage[switch,columnwise]{lineno}

\begin{document}
	\bibliographystyle{apsrev4-2} 
%\linenumbers
\title{Estimation of initial state structures in high energy heavy-ion collisions using Principal Component Analysis (PCA)}
\author{Shreyasi Acharya}
\email{shreyasi.acharya@gmail.com}
\author{Subhasis Chattopadhyay}
\email{sub@vecc.gov.in}
\affiliation{Variable Energy Cyclotron Centre, HBNI, 1/AF, Bidhan Nagar, Kolkata-700064, India}

\begin{abstract}
In high energy heavy ion collisions, structures in the initial collision zone is a matter of intense investigation, both from theory and experimental points of views. A large number of models have been developed to represent the initial state of the collision including Glauber model, Colour Glass Condensate (CGC) among others. Another important aspect of the study is to investigate proper observables that will be sensitive to the initial collision zone. In this work, we have discussed a formalism to implement the spatial clusters at the partonic level in the string melting version of the AMPT model for PbPb collisions at $\sqrt{s_{NN}}$ = 200 GeV. These clusters are then propagated through the AMPT hadronization scheme. The Principal Component Analysis (PCA) has been used on the $\eta$, $\phi$ and $p_T$ distributions of the produced charged particles and the eigenvalues have been compared before and after the implementation of the clustering. It is found that for all these three different distributions, all the prominent PCA modes have shown sensitivity to the clustering. A centrality dependent study has also been performed on those eigenvalues.

\end{abstract}

\maketitle

\section{Introduction} 

In heavy ion collisions at ultra-relativistic energies at RHIC and LHC, corresponding to a medium of high temperature, a state of strongly interacting medium is formed in which partons are de-confined from the incoming hadrons. A range of observables that are measured to characterize the properties of the medium include thermodynamic properties like temperature, entropy, collective properties given by flow parameters, gluon density of the medium measured by jet-quenching among others. It is a usual practice that the experimental observables are theoretically evaluated by folding the space-time evolution of the  colliding medium for finding the sensitivity to the different stages of the evolution\cite{Adams:2005dq, Adcox:2004mh}. {The prominent stages that are modeled include initial state of the collision, formation of the medium, evolution and cooling of the medium, hadronization, rescatterings, chemical and kinetic freeze-out and finally the free streaming of particles.} In the initial stage of the collision, when nucleons overlap, the geometry of the collision zone plays a crucial role in deciding the final state observables{ \cite{Ollitrault:1992bk}.} It has been observed that the final state collectivity parameters commonly known as flow parameters are correlated with the initial state geometry or corresponding fluctuations. {These initial state geometry parameters like various orders of eccentricities from coordinate space ($\epsilon_n$) leave their imprints on the azimuthal distributions of the momentum of the produced particles.} The decomposition of the azimuthal distributions have been represented by parameters of various orders like $v_1, v_2$ etc as obtained by the Fourier decomposition of the azimuthal distributions with respect to the reaction plane angle{\cite{Voloshin:1994mz},} \cite{Heinz:2013th}. The degree of conversion of the initial spatial asymmetry to the final state momentum asymmetry is represented by the correlation between the eccentricities to the flow parameters.  For studying such an effect, one needs to evaluate the event by event eccentricities and hadronic flow parameters. It is however important to investigate in detail the effect of these initial state geometry parameters or their  fluctuations to the distributions of the final-state observables like pseudo-rapidity or transverse momentum distributions of the produced particles. In literature, a range of models describing high energy heavy ion collisions have been discussed that include specific structures of the initial state geometry due to nucleonic overlap or formation of new structures at the partonic or hadronic levels in the form of clusters. Prominent examples include models like Parton cascade Model (PCM)\cite{Muller:2003dn}, Color Glass Condensate (CGC)\cite{Gelis:2010nm}, Zhang parton cascade (ZPC) \cite{Zhang:1997ej} among others. It is a usual practice to implement different initial state scenarios before evolution of the medium using {ideal or viscous hydrodynamics} and a conclusion is made about the suitable description of the initial state that matches the data best. Efforts are also made to study the sensitivity to the fluctuations in the initial state using various methods like sensitivity to the final state observables\cite{Petersen:2010zt, ColemanSmith:2012ka, Floerchinger:2013vua}.  In the present study, we have implemented a clustering algorithm on partons formed by the AMPT model{\cite{Zhang:1999bd}} in PbPb collisions at RHIC energy. The clustering algorithm has been motivated by the formation of spatial domains consisting of thermal partons. These partons are then processed via the hadronization scheme in AMPT string-melting version which is based on the recombination mechanism. The final state particles are then studied in detail. 

In literature,  distributions of the produced particles are analyzed using various decomposition methods like Fourier analysis applied on the azimuthal distributions in order to extract the flow parameters. Recently, Principal Component Analysis (PCA) is being used extensively primarily to study various orders of flow variables and their correlation with the initial geometry parameters and their fluctuations\cite{Bhalerao:2014mua}. There are two main approaches applied in the field of high energy heavy ion collisions for the PCA decomposition. The first one is the decomposition of the covariance of the azimuthal distributions as weighted with a Fourier series and then making connection of the PCA components with the flow parameters \cite{Liu:2019jxg}.  Another approach is to decompose the inclusive distributions using PCA and connect the components with the physical observables like flow parameters in case of decomposition of the azimuthal distributions. Now-a-days PCA is being used extensively in the automated machine learning procedure for finding structures in object spaces. The features primarily detected using PCA are then analysed in details using sophisticated cluster finding algorithm to obtain the features in detail. In our study, we have adopted the later approach i.e. decomposing the inclusive distributions and study the components. {The initial clustering might introduce features in $\eta$, $\phi$, $p_{\rm T}$ space like flow coefficients in azimuthal distribution. As the eigenvalues are sensitive to these features, we can choose a region of eigenvalues that will select the events considered.}

We discuss the AMPT model and the PCA procedure as have been applied in our study in sections II and III respectively. We have then discussed the procedure of implementation of clustering in section-IV, results in section-V and the summary in section-VI.

\section{AMPT}

A Multi-Phase Transport (AMPT) \cite{Zhang:1999bd} is a Monte Carlo partonic transport model being used widely for simulating NN,NA and AA collisions at high energy. The model implements all major stages of the collision starting from the initial state through the partonic scattering followed by hadronization and hadronic rescattering. {The initial stage of the collision has been implemented by HIJING \cite{Gyulassy:1994ew} through Monte-Carlo Glauber model calculations in NA and AA collisions.}  In the initial state of the collisions, either partonic strings and minijets are taken together from HIJING or all strings are melted into partons. There are two versions of AMPT, in the default version, only the minijets are transported using the Zhang's Partonic Cascade (ZPC) and in the string melting version, all melted partons go through ZPC for scattering. The scattering is governed by a parameter to be tuned to match the particle spectra. The hadronization is implemented in two modes known as hadronic mode and partonic mode. In hadronic mode, minijets,  after scattering are recombined with the strings and then get fragmented using Lund's string fragmentation model. On the other hand, in the partonic mode, all the partons combine to form hadrons (mesons or baryons) based on the spatial distance, spin structures and the invariant mass of the quarks (quark-antiquark in case of mesons and 3 quarks in case of baryons). The hadrons formed by any of these two mechanisms then undergo scattering among themselves and then scattered hadrons reach the detector. AMPT has been used extensively in high energy heavy ion collisions and has been able to explain most of the observables like spectra, flow among others. One extremely prominent finding of the model is the ability to explain the {number of constituent quark (NCQ) scaling} of elliptic flow parameter $v_2$ at RHIC. {The NCQ scaling refers to the scaling behavior observed when the $v_2$ and $p_T$ of different identified hadrons are divided by the number of constituent quark ($n_q$). The $v_2/n_q$ vs. $p_T/n_q$ for identified hadrons follow a universal curve suggesting the dominance of quark degrees of freedom at the early stages of collisions.} In our study, we have used only the partonic version of the model involving partons at the initial stage and in hadronization. At the initial stage, a separate partonic clustering has been implemented as discussed in section IV.

\section{Principal Component Analysis (PCA)}

PCA is a method of decomposing a correlated distribution in various components known as principal components that reflect the independent variables characterizing the features of the distributions. PCA is essentially a procedure of dimension reduction from correlated matrix with the eigenvalues representing the variance.

Mathematically, a  matrix (N x m) can be decomposed as

\begin{equation}
 M = X \Sigma Z = VZ
 \end{equation}
where X, Z are orthogonal matrices of N x N and m x m dimensions respectively, {$\Sigma$ is a diagonal matrix of N x m dimensions} with diagonal elements arranged in strict decreasing order. These elements carry physical meaning. In our case, the distribution of a variable in an event can be expressed as 

\begin{equation}
f=\sum_{j=1}^{m}x_j^{(i)} \sigma_j z_j = \sum_{j=1}^{(i)} v_j^{(i)} z_j
\end{equation}

where $z_j$ is an orthogonal vector such that $Z_i ^Tx Z_j = \delta_{ij} $, {$\sigma_j$ are the diagonal elements of matrix $\Sigma$, index i represents the event number (1,2....N) and m is the number of bins of input variable. $v_j^{(i)}$ is the corresponding coefficient of $z_j$ for i$^{th}$ events}. In PCA, $\sigma_j$ are obtained in decreasing order and only a top few values are enough to describe the distribution, say up to $k$, then we can rewrite the equation above as below,

\begin{equation}
f = \sum_{j=1}^{k} v_j^{(i)} z_j
%$f \approx \Sum v_j z_j
\end{equation}
here j is called PCA modes describing the fluctuations in the distribution.

PCA has been used so-far mostly for analysing the covariance of the azimuthal distributions of the produced particles as weighted by a Fourier series primarily to extract the flow coefficients \cite{Mazeliauskas:2015vea, Sirunyan:2017gyb, Bozek:2017thv}. The PCA components represent flow fluctuations in different orders and non-linear couplings among the flow coefficients.
In another approach, however, the inclusive azimuthal distributions are decomposed by PCA and it is found that the eigenvectors of at least up to $4^{th}$ order are similar to the distributions of the Fourier components.  The eigenvectors have been found to be of the shapes similar to that of the Fourier components as has been used in the conventional method of extraction of the flow coefficients. The eigenvalues have been found to correspond to the flow coefficients. If applied at the partonic level, the eigenvalues of PCA correspond to the eccentricities ($\epsilon_n$)  of various orders. {The flow coefficients of various orders have been interpreted to be connected to the initial spatial geometry and their fluctuations to be transferred to the momentum-space anisotropy.} {Flow coefficients (v$_n$) follow a linear relationship with their corresponding initial state eccentricities ($\epsilon_n$)\cite{Alver:2010gr},\cite{Liu:2019jxg}}. In earlier studies using PCA for the azimuthal distributions, event by event $v_n$s have been extracted and then correlated with $\epsilon_n$.
In the present study, event by event distributions for $\eta, \phi, p_T$ have been divided into n-bins separately. The exercise is then undertaken for a large (N) number of events. Such binned distribution per event along with the number of  events form a matrix to be diagonalised. The eigenvalues are obtained in a strictly decreasing order. 

In our study, $\eta, \phi $ and $p_T$ distributions of the produced charged particles are decomposed separately for events having the initial geometry with and without inclusion of additional clustering. The main aim of this study is to investigate the behavior  of the eigenvalues with the changes in the initial conditions. {It has been argued elsewhere that the eigenvalues of PCA correspond to the fluctuations in various orders}{\cite{Ian:2015jc}}. With the modified initial conditions, the fluctuations are expected to change and the PCA eigenvalues should be sensitive to these changes. In conventional approaches, events with high PCA eigenvalues could be extracted and those events could be further investigated using sophisticated cluster-finding method to find the substructures in the set of events. This approach is used in Machine Learning technique quite extensively. {The PCA method in its current form uses the covariance among the data to obtain the results, however the approach is not limited to second order cumulant only. A multivariate cumulants study via their principal components had been first proposed by \cite{Morton:2009llm}, analogous to the usual principal components of a covariance matrix. This cumulant method of principal component analysis has been used in various fields of research such as mathematics, economics and computer science.}

\section{Implementation of clustering}

In the literature, there is a series of models which have implemented the initial states of the high energy heavy ion collisions, a few prominent models that include different initial conditions include NUXUS \cite{Drescher:2000ha} , EPOS\cite{Pierog:2013ria}, MC-KLEIN \cite{Drescher:2006ca}, IP-Glasma \cite{Schenke:2012wb} among others. In the present work, we have implemented clusters at the partonic level which is basically inspired by the discussions on formation of spatial domains at the partonic level.  

We have started with the partons from the AMPT string melting version and implemented the clustering in the following way.
A parton, selected as a seed at random, is taken as the center of a cluster. All partons whose inter-parton distance with respect to the seed parton lie within a certain cluster radius (parameter $R$) are assigned as members of that cluster.
\begin{figure}[h!]
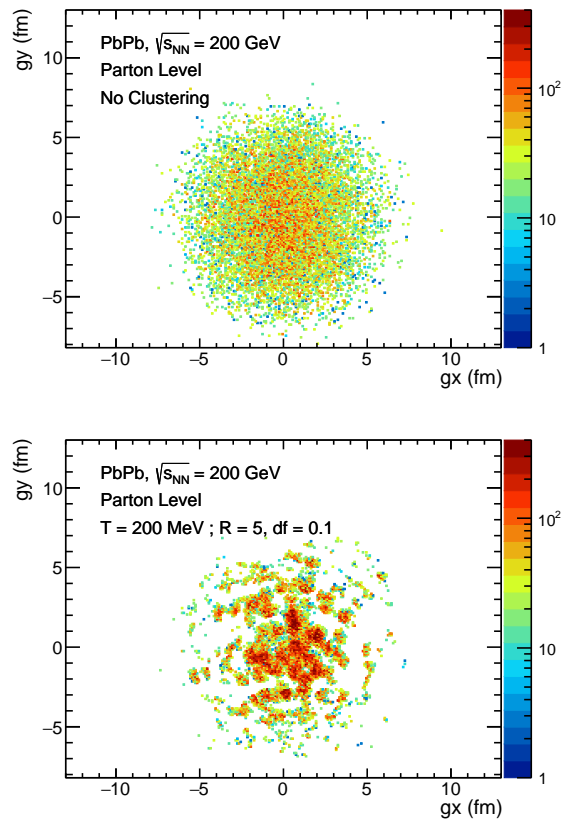

		\includegraphics[scale=0.4]{{2Dgxgy_No_Clustering}.pdf}
	\includegraphics[scale=0.4]{{2Dgxgy_Temp=200MeV_R=5_df=0.1}.pdf}
	
	\caption{X-Y distribution of the partons before (top) and after (bottom) clustering}
	\label{gxgy2D}
\end{figure}
\FloatBarrier The cluster is then formed by bringing the partons closer to the center by reducing the radial distances of the partons by a certain factor (parameter df). Once the formation of one cluster is completed, another unassigned parton is taken as a seed and the process continues till all partons are exhausted.

The next step is to implement a momentum distribution of the members of the partonic cluster. Motivated by the thermally distributed partons, the cluster partons have been assigned momenta as per the following distribution,
\begin{equation}
f(p_T) = e^{{-p_T}/{\rm T}}
\label{eq:MomDist}
\end{equation}
where, T is a parameter with an analogy to the temperature of the cluster. In our study, we have used T values as 200 MeV/c and 400 MeV/c.
The Fig.\ref{gxgy2D} shows the X-Y distribution of the partons on the transverse plane before (top) and after (bottom) clustering. The clustering parameters for the plot are R=5 fm df=0.1, T=200 MeV/c.  As seen in the figure, while before clustering (top), the position distributions of the partons are uniform, clear domain structures are seen in the Fig.\ref{gxgy2D} (bottom) which could be said to correspond to the partonic domains in the position space. Please note that with these parameters, the clusters correspond to {maximum} radii of 0.5 fm {(R$\times$df)}.

\begin{figure}[h!]
 	\includegraphics[scale=0.4]{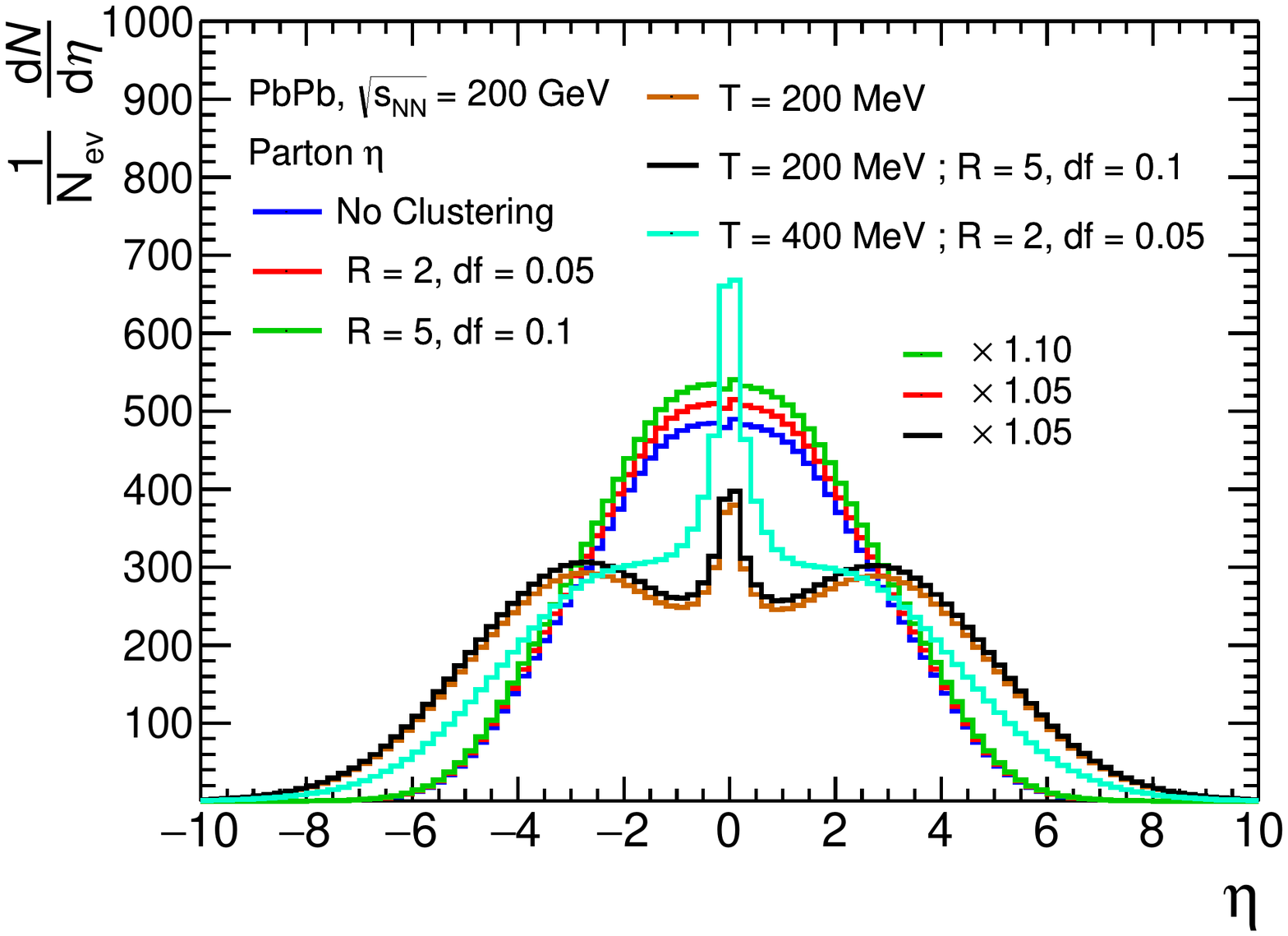}
 	\includegraphics[scale=0.4]{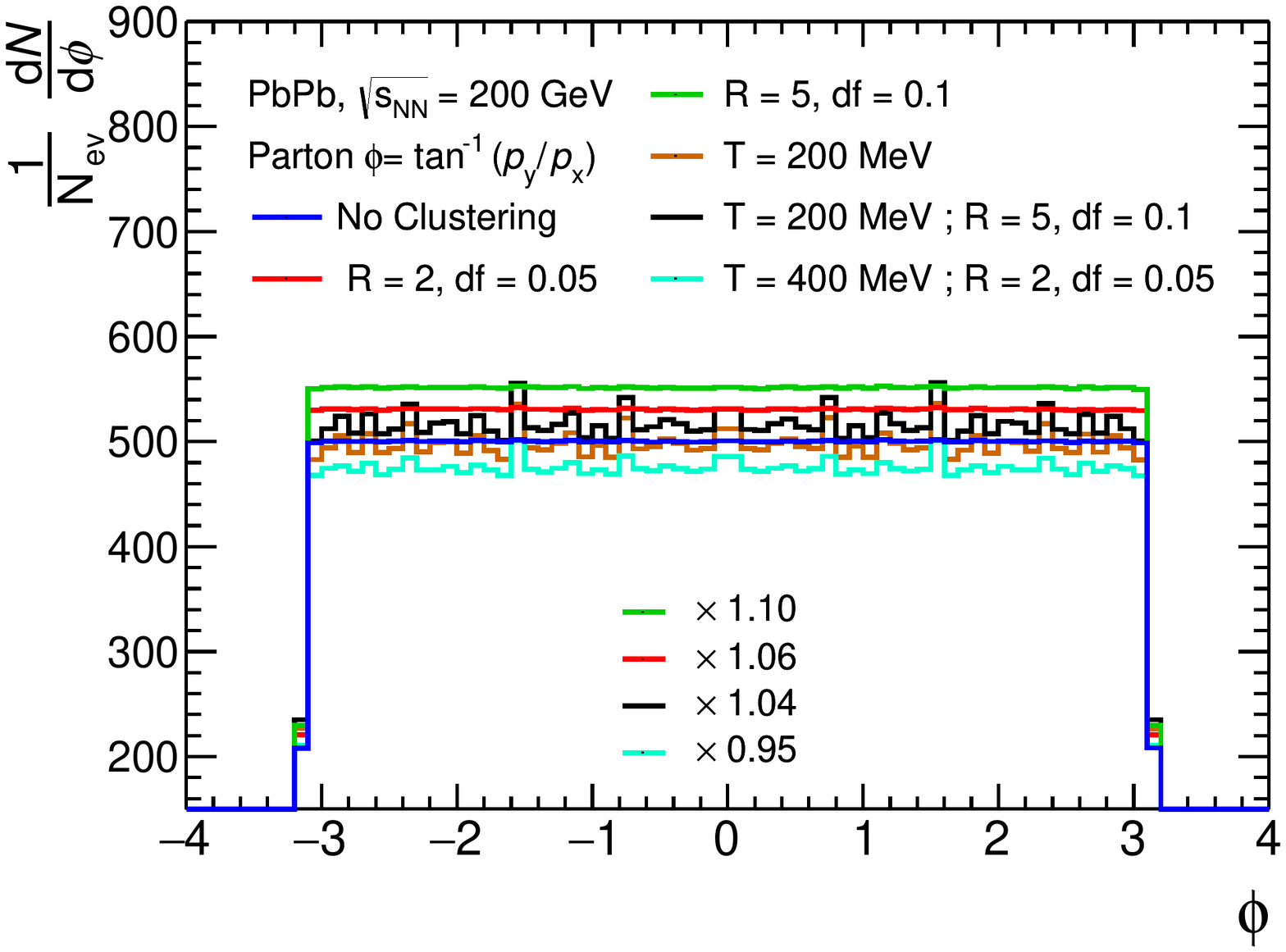}
 	\includegraphics[scale=0.4]{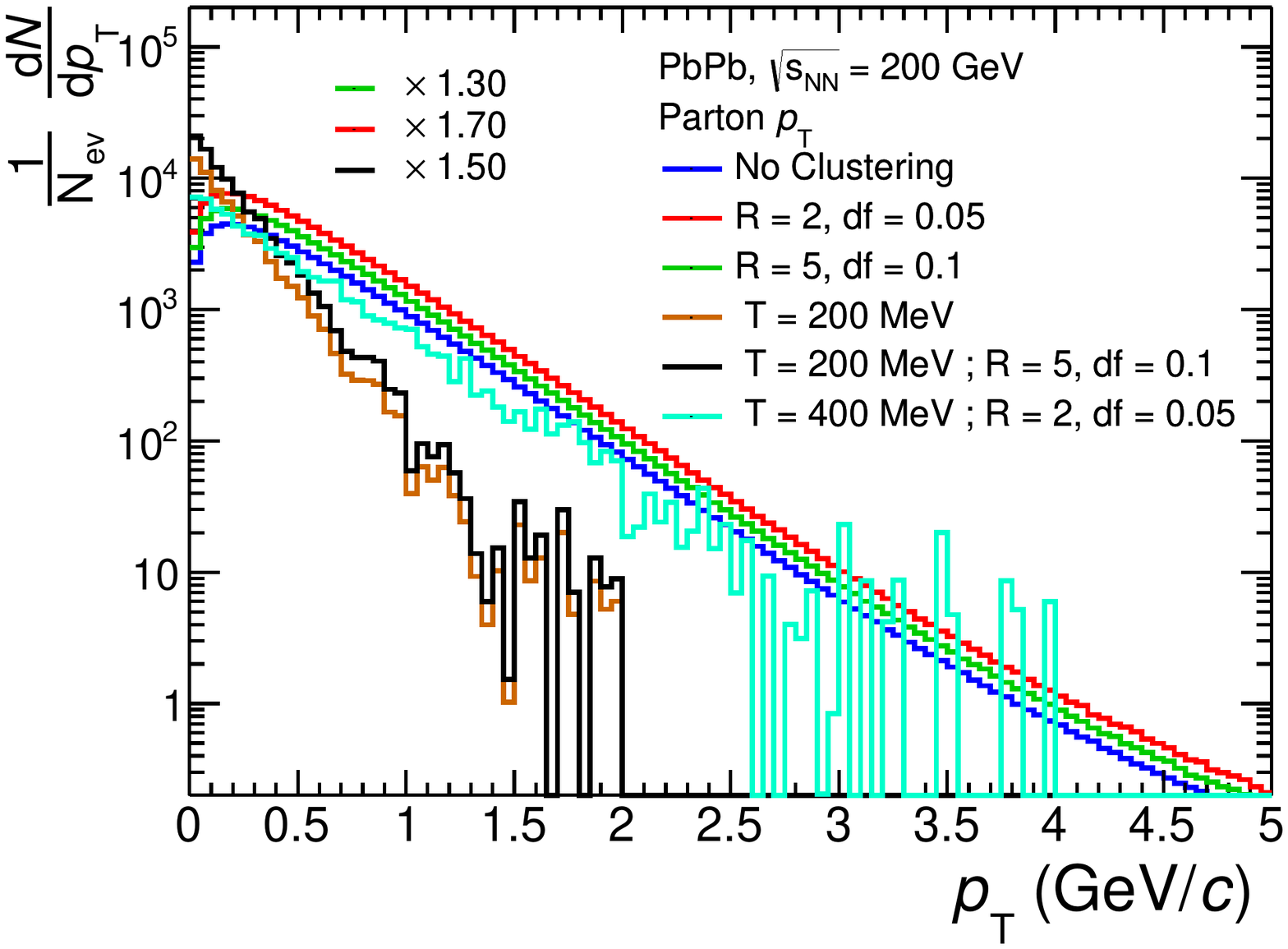}
 	\caption{$\eta, \phi$ and $p_T$ distributions of the partons before and after clustering {for minimum bias Pb-Pb collisions at $\sqrt{s_{NN}}$ = 200 GeV}}
 	\label{PtEtaPhi_Partons}
 \end{figure}

 The Fig.\ref{PtEtaPhi_Partons} shows the $\eta, \phi$ and $p_T$ distributions of the initial partons before and after the clustering with two different temperature parameters and different spatial cluster parameters. {Three cases of clustering has been considered in the figure which are as follows. The legends where only T values (200 MeV or 400 MeV) are mentioned, are cases where the parton momentum has been distributed according to Eq. (\ref{eq:MomDist}) and no position clustering has been implemented. Legends where both T and R, df values are mentioned are the ones where both position clustering has been implemented and parton momentum has been assigned according to Eq. (\ref{eq:MomDist}), and finally the ones with only R, df values and no temperature values are cases where only position clustering has been applied and momentum has not been changed. The $\eta, \phi$ and $p_T$ distributions obtained after applying the aforementioned changes have been compared with the case where no changes in position and momentum has been made.} The  $\eta$ distribution of partons  changes from a uniform to a peaking shape at $\eta=0$ which represents the formation of the clusters. {The $\eta$ distributions for similar temperature parameter overlap e.g., the no-clustering and only position clustering overlap as the momentum has not been changed, while the two curves with T=200 MeV overlap. For better visibility of the plots some of the overlapping curves have been scaled as shown in the legends.}  The azimuthal distributions are mostly uniform except a few hints of azimuthal asymmetry/structure in the clusters which have undergone momentum modification and as expected, $p_T$ distributions depict the modified distribution as per the value of the input T parameter. {Similar to the $\eta$ distributions in the $p_T$ distributions, curves with similar temperature and similar transverse momentum distributions overlap.}
 \begin{figure}[h!]
 	\includegraphics[scale=0.4]{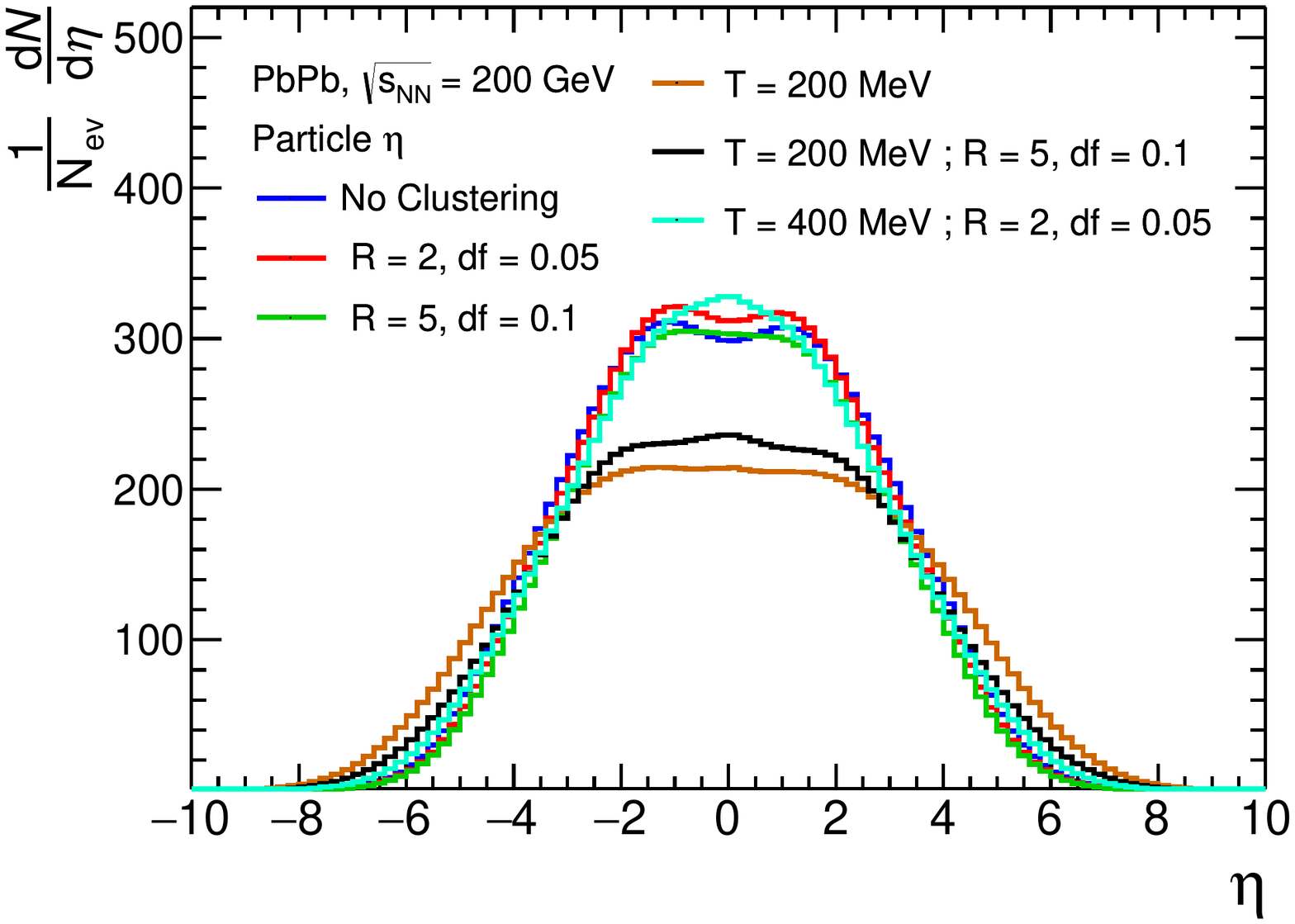}
 	\includegraphics[scale=0.4]{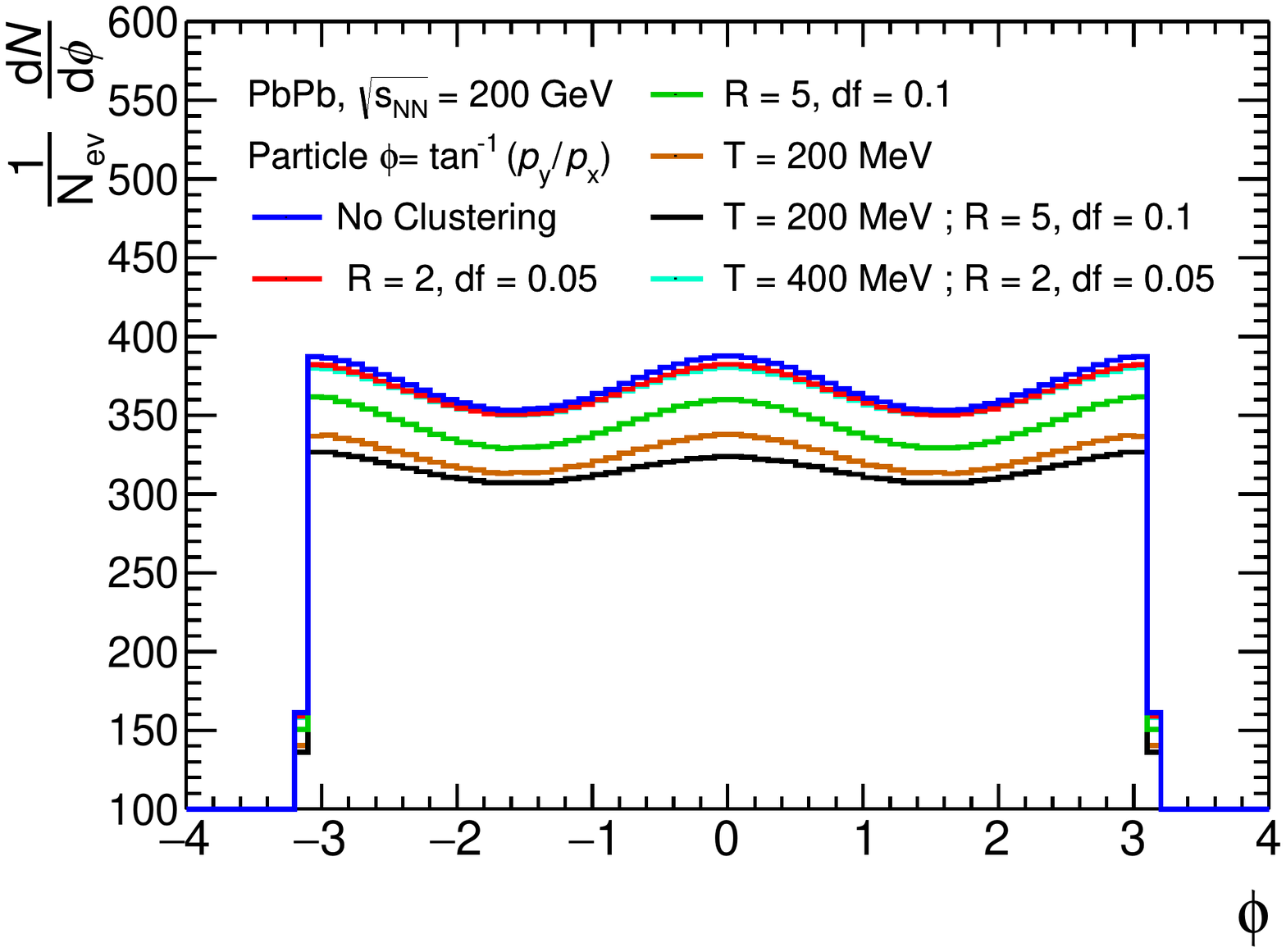}
 	\includegraphics[scale=0.4]{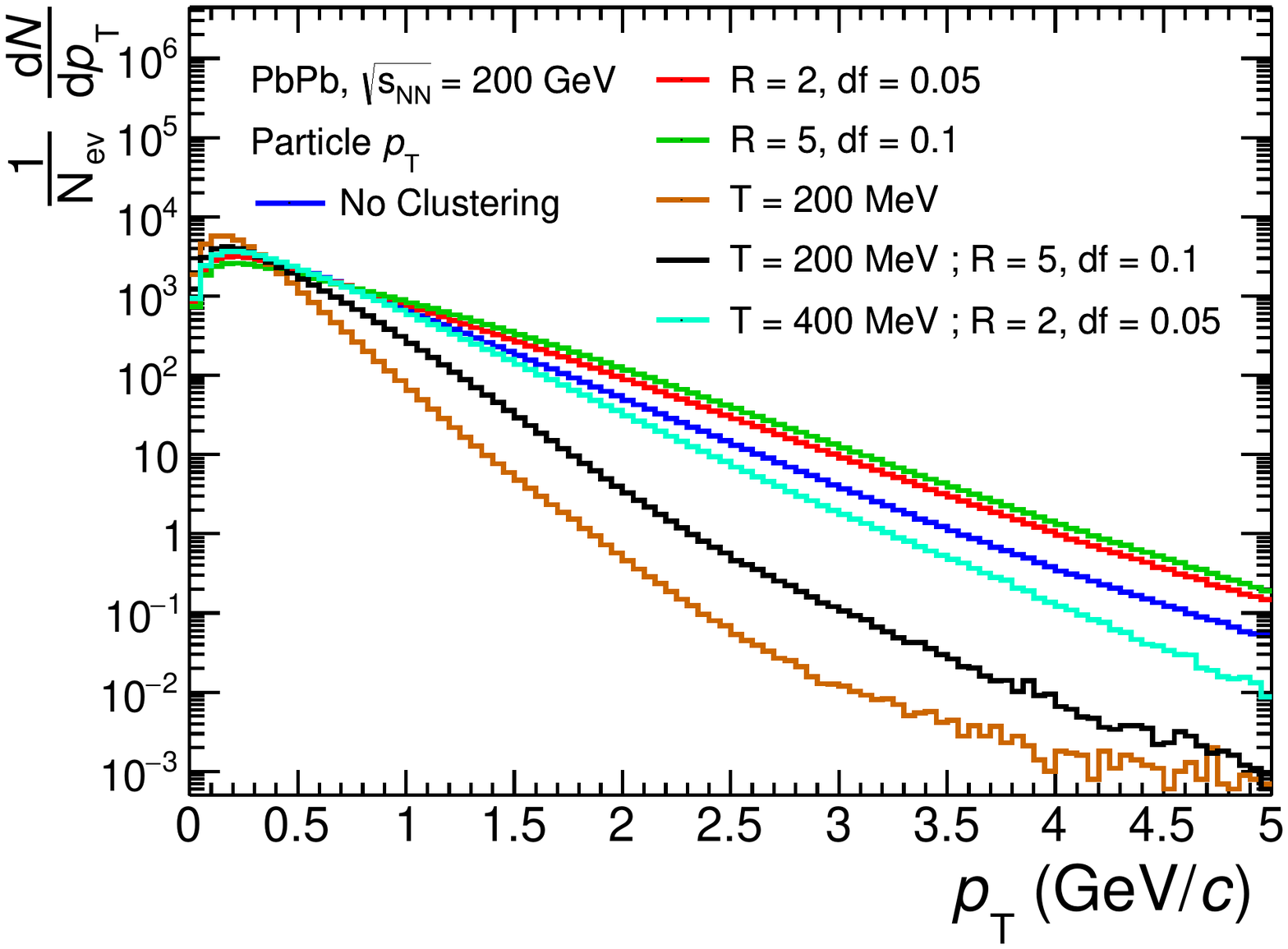}
 	\caption{$\eta, \phi$ and $p_T$ distributions of the produced particles before and after clustering {for minimum bias Pb-Pb collisions at $\sqrt{s_{NN}}$ = 200 GeV}}
 	\label{PtEtaPhi_Particle}
 \end{figure}
 The partons then undergo scattering using ZPC and hadronization as implemented in the partonic version of AMPT i.e., the coalescence of quarks and anti-quarks. We have shown the corresponding distributions of the produced charged particles in the Fig. \ref{PtEtaPhi_Particle}. It is seen that all characteristic structures seen at the partonic level are smoothened out.{The azimuthal distributions of the produced particles on the other hand show the characteristic asymmetric shape due to elliptic flow.} The $p_{\rm T}$ range of the produced particles increases as compared to that of partons due to the production of hadrons consisting of more than one parton. 

% \FloatBarrier
\section{Results}

In this study, we have performed simulation using AMPT string-melting version for Pb+Pb collisions at $\sqrt{s}$ = 200 GeV. We have generated up to {0.2 million} minimum bias events and ensured that the statistical error on the event averaged eigenvalues are not significantly large. We have used only the results from the produced charged particles only in this study.  As discussed earlier in section-III, we divided each event-wise distribution into 20 bins for the $\eta$ and $p_T$ distributions in the regions of -1 to +1 for $\eta$ and 0 to 5 GeV/{\it c} for $p_T$  and 50 bins in the region -$\pi$ to +$\pi$ for $\phi$ . Before discussing the PCA results, we first obtained the elliptic flow parameter $v_2$ using the event plane method{\cite{Poskanzer:1998yz}} for two cases i.e., with and without clustering.  The cluster-parameter have been varied to represent different possibilities.
 In Fig.4 and the subsequent figures, we have opted for two values of the cluster radius parameter (R) i.e., 2 fm and 5 fm associated with parameter df 0.05 and 0.1 respectively. For cases without any mention of the temperature parameters T, parton momenta remain unmodified compared to that from AMPT. For clusters having thermal partons, the T parameters chosen are 200 MeV/c and 400 MeV/c. It might be mentioned that for partons with no clustering, fitted slope of the $p_T$ spectra gives an inverse slope of about 400 MeV/c. The clusters with T = 200 MeV therefore represents significantly softer partons.

\begin{figure}[h!]
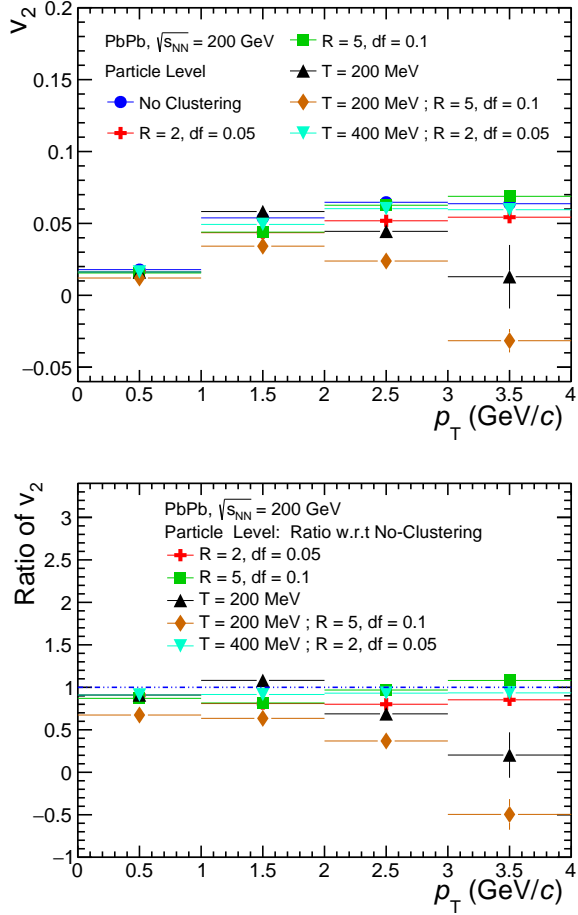

 	\includegraphics[scale=0.4]{{v2_Pt}.pdf}
 	\includegraphics[scale=0.4]{{Ratio_v2_Pt}.pdf}
 	\caption{$p_T$ distributions of $v_2$ for two scenarios i.e., before and after clustering (upper panel) and the ratio of $v_2$ vs. $p_{T}$ distributions after clustering w.r.t the one before. }
 	\label{v2_pt}
 \end{figure}

The Fig. \ref{v2_pt} (top) shows the variation of $v_2$ with $p_T$ for two scenarios. As seen in the Fig.\ref{v2_pt} (top), $v_2$ increases with $p_T$ except for T= 200 MeV in which $v_2$ reduces at higher $p_T$. 
We have also shown the ratio of $v_2$ in the Fig. \ref{v2_pt} (bottom) taking no-clustering scenario as reference and as discussed earlier, the ratio remains constant at unity thereby insensitive to the clustering except the T = 200 MeV case.  

\begin{figure}[h!]
	\includegraphics[scale=0.4]{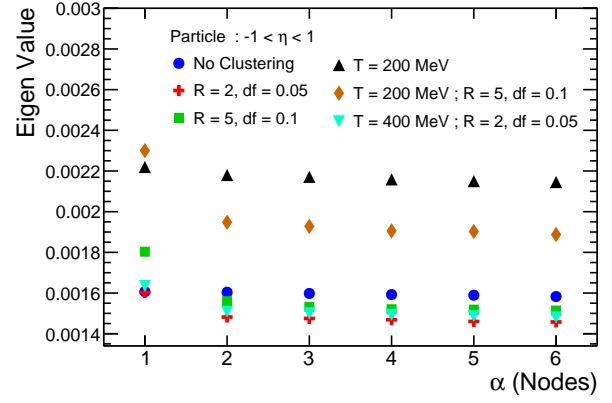}
	\caption{Eigenvalues as obtained for $\eta$ distribution before and after clustering}
	\label{EigenVal_Eta}
\end{figure}

\begin{figure}[h!]
	\includegraphics[scale=0.4]{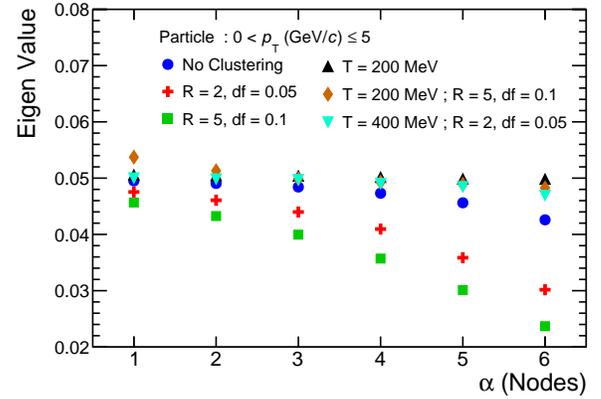}
	\caption{Eigenvalues as obtained for $p_T$ distribution before and after clustering}
	\label{EigenVal_Pt}
\end{figure}

 %\FloatBarrier
As mentioned earlier, the eigenvalues of PCA  are related to the eccentricities at the partonic level and various flow components and their fluctuations as obtained from the azimuthal distributions at the particle level. We have investigated the eigenvalues for different distributions  of the charged particles with different cluster parameters.
The Fig. \ref{EigenVal_Eta} to Fig. \ref{EigenVal_Pt} show the distributions of the eigenvalues as obtained for $\eta$, $\phi$ and $p_T$ distributions for two different scenarios i.e. with and without clustering. The nodes ($\alpha$) in the X-axis represents the PCA components. It is clear in all cases that the most prominent eigenvalue is of the first component differing from the next one by varying degree. It can therefore be mentioned that the first eigenvalue, representing the variance of the distribution of the reduced dimension can be used further  for investigating the structures in the initial state of the collision zone.

 As per the PCA method, eigenvalues are arranged in a decreasing order sometime with a wide difference between the eigenvalue of component-1 to that of the next node. Before discussing PCA results, we might have a re-look at the Fig. \ref{PtEtaPhi_Partons} and the Fig. \ref{PtEtaPhi_Particle} showing the inclusive distributions of partons and of the produced charged particles respectively. It is clearly seen that partonic $\eta$ and $\phi$ distributions have structures with more prominent ones for the $\eta$ distributions presumably due to the inclusion of clustering at the partonic level. However, at the particle level, no such structures are prominently visible. In view of this, it is important to study the PCA-eigenvalues at the particle level with and without clustering. 
  
 For the $p_{\rm T}$ distribution, the position clustering lowers the eigenvalues compared to the no-clustering case while the clusters including thermal partons tend to increase the eigenvalues for all nodes.  The same pattern is also seen in the $\eta$ distributions,  with exception to the first node ($\alpha$=1), where  for cases involving clustering with R= 5 fm have a higher  eigenvalue than the no-clustering scenario.
 {In case of the $\phi$ distributions, eigenvalues are seen closer in pairs presumably representing the real and imaginary components of the flow parameters}{\cite{Liu:2019jxg}}. We have not made any detailed investigation towards extraction of flow parameters from these eigenvalues. We have only pointed out that the eigenvalues differ clearly for two cases i.e., with and without clustering. We have also observed a clearer effect of the position clustering in case of $\phi$ as for eigenvalues for R= 5 fm lie considerably higher compared to the no-clustering values. The eigenvalues of the $\phi$ distributions look more sensitive to the position clustering. 
 
\FloatBarrier
\FloatBarrier
\begin{figure}[h!]
	\includegraphics[scale=0.4]{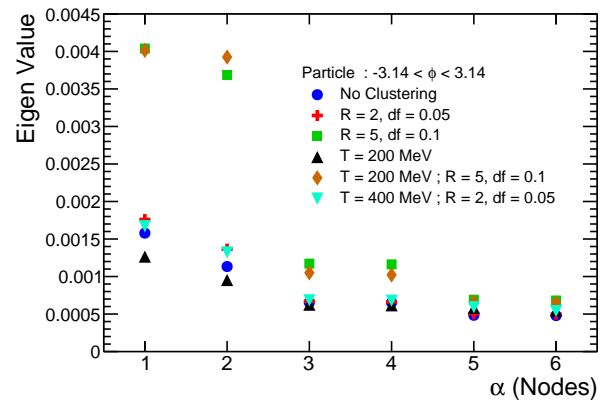}
	\caption{Eigenvalues as obtained for  $\phi$ distribution before and after clustering}
	\label{EigenVal_Phi}
\end{figure}

\begin{figure}[h!]
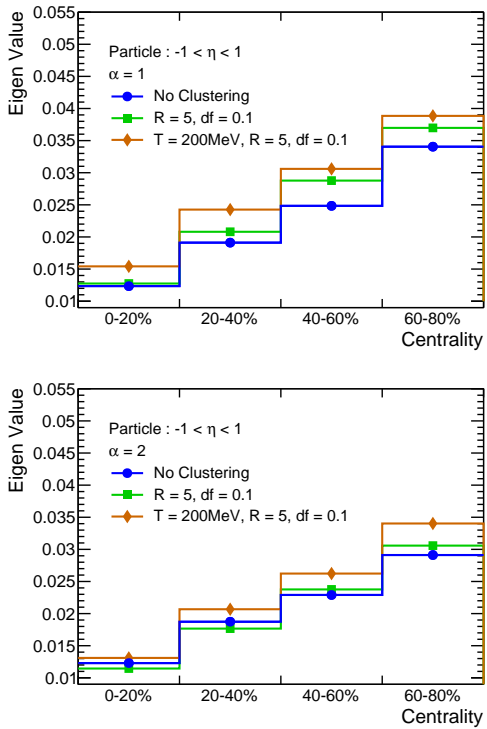

	\includegraphics[scale=0.35]{{yCentrality_vs_EgnVal_3variations_Eta_node=1}.pdf}
	\includegraphics[scale=0.35]{{yCentrality_vs_EgnVal_3variations_Eta_node=2}.pdf}
	\caption{Eigenvalues as obtained for $\eta$ distribution before and  after clustering}
	\label{EigenVal_CentClass_Eta}
\end{figure}

We have also performed studies for various event centralities. The Fig. \ref{EigenVal_CentClass_Eta} to Fig. \ref{EigenVal_CentClass_Pt} show the variation of the eigenvalues of two different modes (1-2) with event centralities.
We have only taken two cases that showed maximum effect in eigenvalue studies shown earlier i.e., (i)  R=5fm, {\it df} = 0.1 and no momentum modifications  and (ii) T=200 MeV, R= 5fm and {\it df} = 0.1. 
It is seen that the eigenvalues of the first component for the $\phi$ and $p_T$ distributions have a decreasing trend for the events where clustering is implemented as compared to the events without clustering.

The observed decreasing trend of the first eigenvalues might be due to higher fluctuations for lower multiplicities in peripheral events. No significant structures are seen for the $\eta$ distributions of the produced particles in both the cases. 
It is also seen that the eigenvalues are considerably lower in case minimum-bias events as shown in the Fig.4-6 discussed earlier. This might be due to dilution of fluctuations for minimum-bias events due to the admixture of events with different multiplicities. 

%\FloatBarrier
\begin{figure}[h!]
	\includegraphics[scale=0.35]{{yCentrality_vs_EgnVal_3variations_Phi_node=1}.pdf}
	\includegraphics[scale=0.35]{{yCentrality_vs_EgnVal_3variations_Phi_node=2}.pdf}
	\caption{Eigenvalues as obtained for  $\phi$ distribution before and  after clustering}
	\label{EigenVal_CentClass_Phi}
\end{figure}
%\FloatBarrier
\begin{figure}[h!]
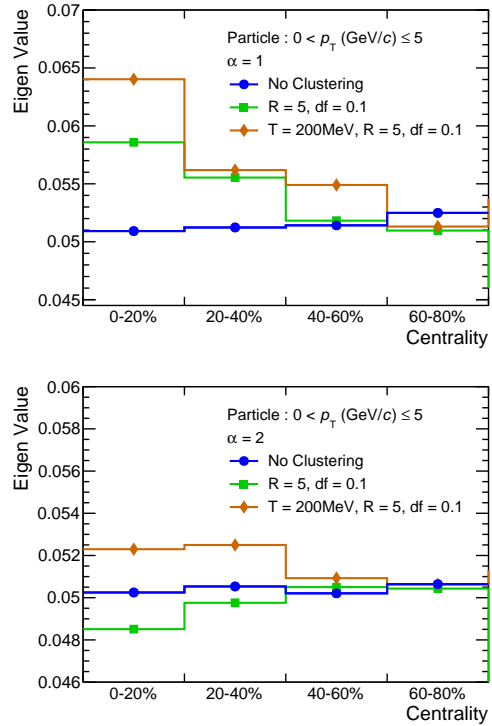

	\includegraphics[scale=0.35]{{yCentrality_vs_EgnVal_3variations_Pt_node=1}.pdf}
	\includegraphics[scale=0.35]{{yCentrality_vs_EgnVal_3variations_Pt_node=2}.pdf}
	\caption{Eigenvalues as obtained for $p_T$ distribution before and after clustering}
	\label{EigenVal_CentClass_Pt}
\end{figure}
%\FloatBarrier  

\section{Summary and Conclusions}

In an effort to find a method to investigate the initial partonic structure in high energy heavy collisions, we have implemented the formation of partonic clusters using the partons obtained from the AMPT model of string-melting version.
The clusters are formed in two steps, first by bringing partons closer in positions to an extent defined by two parameters i.e., the radius of the partonic zone (R) and the scaling factor on the inter-partonic distance. In our work, we have used values as R= 2 fm, 5m and df = 0.05 and 0.1. 
Additionally, we have introduced a thermal distribution to the cluster partons by tuning the temperature parameters, we have used two temperature values i.e., T=200 MeV and T=400 MeV. The later one is close to the inverse slope of the $p_{\rm T}$ distribution from the AMPT partons.
These partons then undergo hadronization by AMPT string-melting hadronization scheme i.e, by coalescence of partons as per their distance, spin and mass. 
We have then made investigation on the distributions of the produced particles from AMPT in order to find the sensitivity of the particle-level observables to the partonic structures. Even though the structures are reflected in basic distributions of the partons, however, there is no clue of these structures in the inclusive distributions of the produced particles. For this investigation, we have used the Principal Component Analysis (PCA) to analyse the $\eta$, $\phi$ and $p_T$ distributions of the produced particles. {It may be mentioned that the square-root of the sum of the squares of the paired eigenvalues from the azimuthal distribution of a particular order has been shown to be related to the coefficients of flow up to v$_6$ \cite{Liu:2019jxg}. In our work we have taken the eigenvalues as our candidate for probing the initial state at different clustering conditions.} For our study, we have looked into the eigenvalues obtained from PCA decomposition {of $\eta$, $\phi$ and $p_T$ distributions} as our observables and looked at them for various conditions like no clustering, only position clustering, inclusion of thermal partons with T= 200 MeV and 400 MeV.  It is found that the first few prominent eigenvalues for all three distributions are sensitive to the inclusion of clustering. For $\eta$ and $p_T$ distributions, two clear groups are seen lying above and below the no-clustering scenario. For T= 200 MeV, all eigenvalues lie above the no-clustering reference. For position clustering, the eigenvalues are grouped below the reference. It is seen that the difference with the no-clustering reference is more for higher values of the R parameter. For the azimuthal distributions, the eigenvalues of which are related to the flow parameters, it appears the the sensitivity is higher towards the position clustering. We have also studied the centrality dependence of the first two eigenvalues. Even though the $\eta$-values do not show appreciable sensitivity, for $\phi$ and $p_T$, they show clearly different trend as compared to the no-clustering reference, which is mostly flat.
We therefore conclude that the first few eigenvalues are sensitive to the inclusion of domains at the partonic level. The events with domains might be identified on an event by event basis by discriminating based on the eigenvalues. 
It is already known that the eigenvalues of the azimuthal distributions represent the flow parameters. In general the PCA eigenvalues represent fluctuations in the distributions of different orders, which are not visible in the inclusive distributions, however, further analysis using the PCA might be performed to extract the physical interpretations of the eigenvalues and eigenvectors from the $\eta$ and $p_T$ distributions.

\section{Acknowledgement}
We would like to thank the VECC grid computing facility for helping in performing the computing for this work. We also acknowledge the funding from the Department of Atomic Energy, Govt of India.

\bibliography{biblio}
\end{document}